\documentclass{article}
\usepackage{amsmath}
\usepackage{icrctc07}

\newcommand{\bem}{\begin{pmatrix}}
\newcommand{\eem}{\end{pmatrix}}
\newcommand{\be}{\begin{equation}}
\newcommand{\ee}{\end{equation}}
\newcommand{\bbe}{\begin{equation*}}
\newcommand{\eee}{\end{equation*}}
\newcommand{\mbf}[1]{\mathbf{#1}}
\newcommand{\mr}[1]{\mathrm{#1}}
\newcommand{\bea}{\begin{eqnarray}}
\newcommand{\eea}{\end{eqnarray}}
\newcommand{\bbea}{\begin{eqnarray*}}
\newcommand{\eeea}{\end{eqnarray*}}
\newcommand{\dEdX}{w}
\newcommand{\Xmax}{\ensuremath X_\mathrm{max}}

\newcommand{\Ne}{\ensuremath N^\mathrm{e}}
\def\EeV{\ifmmode {\mathrm{\ Ee\kern -0.1em V}}\else
                   \textrm{Ee\kern -0.1em V}\fi}%
\def\eV{\ifmmode {\mathrm{\ e\kern -0.1em V}}\else
                   \textrm{e\kern -0.1em V}\fi}%

\title{Longitudinal Shower Profile Reconstruction from Fluorescence and 
Cherenkov Light}
\shorttitle{Longitudinal Shower Profile Reconstruction from Fluorescence and 
Cherenkov Light}

\authors{M.\ Unger, R.\ Engel, F.\ Sch\"ussler and R.\ Ulrich}
\shortauthors{M.\ Unger, R.\ Engel, F.\ Sch\"ussler and R.\ Ulrich}
\afiliations{Forschungszentrum Karlsruhe, Postfach 3640, 76021 Karlsruhe, Germany}
\email{Michael.Unger@ik.fzk.de}

\abstract{Traditionally, longitudinal shower profiles are reconstructed in
fluorescence light experiments by treating the Cherenkov light
contribution as background.  Here we will argue that, due to
universality of the energy spectra of electrons and positrons, both
fluorescence and Cherenkov light can be used simultaneously as signal
to infer the longitudinal shower development.  
We present a new profile reconstruction method that is based on the analytic
least-square solution for the estimation of the shower profile from the observed
light signal and discuss the extrapolation of the profile
with a Gaisser-Hillas function.}
\begin{document}
\maketitle
\section{Introduction}
During its passage through the atmosphere of the earth an extensive air shower
excites nitrogen molecules of the air, which subsequently radiate isotropically 
ultraviolet fluorescence light. Since the amount of emitted light is 
proportional to the energy deposited, the longitudinal shower development
can be observed by appropriate optical detectors such as HiRes~\cite{fd:HiRes}, 
Auger~\cite{fd:Auger} or TA~\cite{fd:TA}.\\
As part of the charged shower particles travel faster than the speed of light in air,
Cherenkov light is emitted in addition. Therefore, in general a mixture of the
two light sources reaches the aperture of the detector.\\ 
In the traditional method \cite{Baltrusaitis:1985mx}
for the reconstruction of the longitudinal shower development
the Cherenkov light is iteratively subtracted from 
the measured total light. The drawbacks of this ansatz are 
the lack of convergence for events with a large amount of 
Cherenkov light and the difficulty of propagating the
uncertainty of the subtracted signal to the reconstructed shower profile.\\
It has already been noted in~\cite{cher:giller} that, due to the universality
of the energy spectra of the secondary electrons and positrons within an
air shower, there exists a non-iterative solution for the reconstruction
of a longitudinal shower profile from light detected by fluorescence telescopes.\\
Here we will present the analytic
least-square solution for the estimation of the shower profile from the observed
light signal in which both, fluorescence and Cherenkov light,
are treated as signal. 
\begin{figure*}
 \centering
 \includegraphics[clip,width=0.86\textwidth]{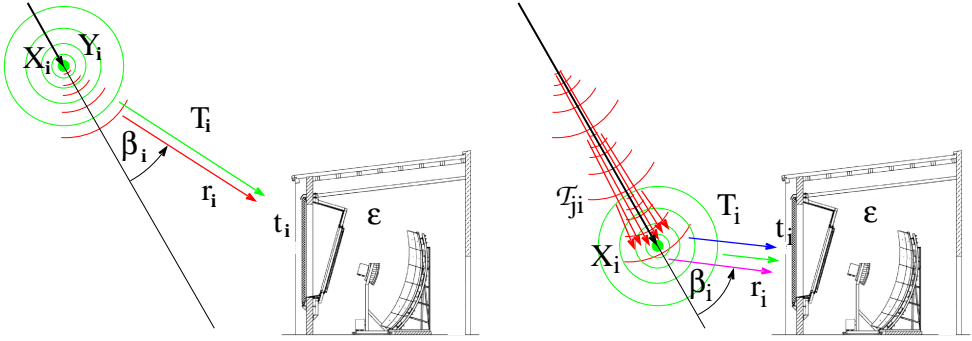}
 \caption{Illustration of the isotropic fluorescence light
          emission (circles), Cherenkov beam along the shower axis and the 
          direct (left) and scattered (right) Cherenkov light contributions.}\label{fig1}
\end{figure*}
\section{Scattered and Direct Light}
The non-scattered, i.e. direct fluorescence light emitted at a certain slant depth $X_i$ is measured
at the detector at a time $t_i$. Given the fluorescence yield $Y^\mr{f}_i$ 
\cite{Nagano:2004am, Kakimoto:1995pr} 
at this point of the atmosphere, the number of photons produced at the shower in a slant
depth interval $\Delta X_i$ is 
\bbe
N_\gamma^\mr{f}(X_i) =  Y^\mr{f}_i\,\dEdX_i\,\Delta X_i,
\eee
where $\dEdX_i$ denotes the energy deposited at slant depth $X_i$ (cf.\ Fig.~\ref{fig1}). 
These photons are distributed over
a sphere with surface $4\,\pi\,r_i^2$, where $r_i$ denotes the distance of the detector.
Due to atmospheric attenuation only a fraction $T_i$ of them can be detected. Given
 a light detection efficiency of $\varepsilon$, the measured fluorescence light flux 
$y_i^\mr{f}$ can be written as
\be
  y_i^\mr{f}=  d_i\,Y^\mr{f}_i \,\dEdX_i\,\Delta X_i,
  \label{eq:fluo}
\ee
where the abbreviation
$
  d_i=\frac{\varepsilon\, T_i }{4\,\pi\,r_i^2} 
$ 
was used. For the sake of clarity
the wave length dependence of $Y$, $T$ and $\varepsilon$ 
will be disregarded in the following but be discussed later.\\
The number of Cherenkov photons emitted at the shower is proportional to the 
number of charged particles above the Cherenkov threshold energy. Since the 
electromagnetic component dominates the shower development, the emitted Cherenkov light, 
$N_\gamma^\mr{C}$, can e calculated from 
\bbe
N_\gamma^\mr{C}(X_i) = Y^\mr{C}_i\,\Ne_i\,\Delta X_i,
\eee
where $\Ne_i$ denotes the number of electrons and positrons above a certain energy cutoff, which
is constant over the full shower track and not to be confused with the Cherenkov emission
energy threshold. Details of the Cherenkov light
production like these thresholds are included
in the Cherenkov yield factor $Y^\mr{C}_i$ \cite{cher:giller,cher:hillasangle,cher:hillaslongi,cher:nerling}. 

Although 
the Cherenkov photons are emitted in a narrow cone along the particle direction, they 
cover a considerable angular range with respect to the shower axis, because
the charged particles are deflected from the primary particle direction due to multiple
scattering. Given the fraction $f_\mr{C}(\beta_i)$ of Cherenkov photons emitted at an angle 
$\beta_i$ with respect
to the shower axis~\cite{cher:hillasangle,cher:nerling}, the light flux at the detector 
aperture originating from direct Cherenkov light is
\be
  y_i^\mr{Cd}= d_i\, f_\mr{C}(\beta_i)\,Y^\mr{C}_i\,\Delta X_i \, \Ne_i.
  \label{eq:dcher}
\ee
Due to the forward peaked nature of Cherenkov light production, an intense Cherenkov
light beam can build up along the shower as it traverses the atmosphere (cf.~Fig.~\ref{fig1}).
If a fraction $f_\mr{s}(\beta_i)$ of the beam is scattered towards the detector it can
contribute significantly to the total light received. In a simple one-dimensional model
the number of photons in the beam at depth~$X_i$ is just the sum of Cherenkov light produced at 
all previous depths $X_j$ attenuated on the way from $X_j$ to $X_i$ by $\mathcal{T}_{ji}$:
\bbe
N_\gamma^\mr{beam}(X_i) = \sum_{j=0}^i \mathcal{T}_{ji}\,Y^\mr{C}_j\,\Delta X_j\,\Ne_j.
\eee
Similar to the direct contributions, the scattered Cherenkov light received at the detector is then
\be
  y_i^\mr{Cs}= d_i\, f_\mr{s}(\beta_i)\,\sum_{j=0}^i \mathcal{T}_{ji}\,Y^\mr{C}_j\,\Delta X_j\,\Ne_j.
  \label{eq:scher}
\ee
Finally, the total light received at the detector at the time $t_i$ is obtained by 
adding the scattered and direct
light contributions.
\section{Shower Profile Reconstruction}
\label{sec:reco}
The aim of the profile reconstruction is to estimate the energy deposit and/or 
electron profile from the light flux observed at the detector. At
first glance this seems to be hopeless, since at each depth there are the
two unknown variables $\dEdX_i$ and $\Ne_i$, and only one measured quantity,
namely $y_i$.
Since the total energy deposit is just the sum of the energy loss of electrons,
$\dEdX_i$ and $\Ne_i$ are related via
\be
  \dEdX_i=\Ne_i \, \int_0^\infty f_\mr{e}(E,X_i) \;\dEdX_\mr{e}(E)\; \mr{d}E,
  \label{eq:nededx}
\ee
where $f_\mr{e}(E,X_i)$ denotes the normalized electron energy distribution 
and $\dEdX_\mr{e}(E,X_i)$ is the energy loss of a single electron with energy $E$.
As it is shown in~\cite{cher:hillaslongi,cher:giller,cher:nerling}, the electron energy 
spectrum $f_\mr{e}(E,X_i)$ is universal in shower age $s_i=3/(1+2\Xmax/X_i)$,
 i.e.~it does not depend on the primary mass or energy, but only on
the relative distance to the shower maximum, $\Xmax$. Eq.~(\ref{eq:nededx}) can
thus be simplified to
\bbe
   \dEdX_i=\Ne_i \; \alpha_i.
   \label{eq:dne}
\eee
where $\alpha_i$ is the average energy deposit per electron at shower age 
$s_i$. 
 With this one-to-one 
relation between the energy deposit and the number of electrons, the shower profile
is readily calculable from the equations given in the last section.
For the solution of the problem, it is convenient to rewrite the 
relation between energy deposit and
light at the detector
in matrix notation:
Let 
$
\mbf{y} = 
( y_1, y_2, \dots, y_n )^\mr{T}
$ 
 be the $n$-component vector (histogram) of the measured photon flux  at the aperture and
$
\mbf{\dEdX} = 
( \dEdX_1, \dEdX_2, \dots, \dEdX_n )^\mr{T}
$ 
the energy deposit vector at the shower track. Using the ansatz
\be
   \mbf{y}=\mbf{C}\cdot\mbf{\dEdX}
   \label{eq:matrix}
\ee
the elements of the {\itshape Cherenkov-fluorescence matrix} $\mbf{C}$ can be found
by a comparison with the coefficients in equations~(\ref{eq:fluo}), (\ref{eq:dcher}) 
and (\ref{eq:scher}):
\be
C_{ij}=
\begin{cases}
   0,\hfill i<j\!\;\\
   c_{i}^\mr{d}+c_{ii}^\mr{s},\;\;\; i=j\,\\
   c_{ij}^\mr{s},\hfill i>j,\\
\end{cases}
\label{eq:elements}
\ee
where
\bbe
c_{i}^\mr{d}=d_i \left(Y^\mr{f}_i + f_\mr{C}(\beta_i)\,Y^\mr{C}_i/\alpha_i\right)\,\Delta X_i
\eee
and
\bbe
c_{ij}^\mr{s}=d_i\,  f_\mr{s}(\beta_i)\,\mathcal{T}_{ji}\,Y^\mr{C}_j/\alpha_j\,\Delta X_j .
\eee
The solution of Eq.~(\ref{eq:matrix}) can be obtained by inversion, leading to the energy deposit
estimator $\widehat{\mbf{\dEdX}}$:
\bbe
    \widehat{\mbf{\dEdX}}=\mbf{C}^{-1}\cdot\mbf{y}\,.
\eee
Due to the triangular structure of the Cherenkov-fluorescence matrix the inverse can
be calculated fast even for matrices with large dimension. As the matrix elements in (\ref{eq:elements})
are always $\ge 0$, $\mbf{C}$ is never singular.\\
The statistical uncertainties of $\widehat{\mbf{\dEdX}}$ are obtained by error propagation:
\bbe
\mbf{V_\dEdX}=\mbf{C}^{-1} \,\mbf{V_y} \left(\mbf{C}^{\mr{T}}\right)^{-1}\;.
\eee  
It is interesting to note that even if the measurements $y_i$ are uncorrelated, i.e.~their 
covariance matrix $\mbf{V_y}$ is diagonal, the calculated energy loss values 
$\widehat{\dEdX}_i$ are not. This is, 
because the light observed during time interval
$i$ does not solely originate from $\dEdX_i$, but  also receives 
a contribution from earlier shower parts $\dEdX_j$, $j<i$, 
via the 'Cherenkov beam'.\\
\section{Wavelength Dependence}
\label{sec:lambda}
Until now it has been assumed that the shower induces light emission at a single wavelength $\lambda$. 
In reality, the fluorescence yield
shows distinct emission peaks and the number of Cherenkov photons is proportional
to $\frac{1}{\lambda^2}$. In that case, also the wavelength dependence of the
detector efficiency and the light transmission need to be taken into account.
Assuming that a binned wavelength distribution of the yields is available
($
   Y_{ik}=\int_{\lambda_k-\Delta \lambda}^{\lambda_k+\Delta \lambda} Y_{i}(\lambda)\, \mr{d}\lambda
$), 
the above considerations still hold when replacing $c_i^\mr{d}$ and
$c_{ij}^\mr{s}$ in Eq.~(\ref{eq:elements}) by
\bbe
\tilde{c}_i^\mr{\,d}=\Delta X_i\sum_k\,d_{ik} \left(Y^\mr{f}_{ik} +
                     f_\mr{C}(\beta_i)\,Y^\mr{C}_{ik}/\alpha_i\right)
\eee
and
\bbe
\tilde{c}_{ij}^\mr{\,s}=\Delta X_j\sum_k\,d_{ik}\, f_\mr{s}(\beta_i)\,
                        \mathcal{T}_{jik}\,Y^\mr{C}_{jk}/\alpha_j,
\eee
where
\bbe
  d_{ik}=\frac{\varepsilon_k\, T_{ik} }{4\,\pi\,r_i^2}.
\eee
The detector efficiency $\varepsilon_k$ and transmission coefficients $T_{ik}$ 
and $\mathcal{T}_{jik}$ are evaluated at the wavelength~$\lambda_k$.
\section{Shower Age Dependence}
Due to the age dependence of the electron spectra $f_\mr{e}(E,s_i)$, the
Cherenkov yield  factors $Y^\mr{C}_i$ and the average electron energy deposits $\alpha_i$
depend on the shower maximum, which is not known before the profile has been reconstructed. 
Fortunately, these dependencies are small: In the age range of importance for the
shower profile reconstruction ($s\in[0.8,1.2]$) $\alpha$ varies only within
a few percent \cite{cher:nerling} and $Y^\mr{C}$ by less than 15\%~\cite{cher:giller}.
Therefore, a good estimate of $\alpha$ and $Y^\mr{C}$ can be obtained by setting
$s=1$. After the shower profile has been calculated with these estimates, 
 $\Xmax$ can be determined and the profiles can be re-calculated with an updated
Cherenkov-fluorescence matrix.\\

\section{Gaisser-Hillas Fit}
The knowledge of the complete profile is required for the calculation of the Cherenkov beam 
and the shower energy.
If due to the limited field of view of the detector only a part of the profile is observed, 
an appropriate function for the extrapolation to unobserved depths is needed. 
A possible choice is the Gaisser-Hillas 
function~\cite{ghfunc}
which was found to give a good description of measured longitudinal profiles~\cite{ghhires}. 
It has only four free parameters: 
$\Xmax$, the depth where the shower reaches its maximum energy deposit $\dEdX_\mr{max}$
and two shape parameters $X_0$ and $\lambda$.\\
The best
set of Gaisser-Hillas parameters $\mbf{p}$ can be obtained by minimizing
the error weighted squared difference between the vector of function
values $\mbf{f_\mr{GH}}$ and $\widehat{\mbf{x}}$, which is
\bbe
 \chi^2_\mr{GH} = \left[\,\widehat{\mbf{\dEdX}}-\mbf{f(\mbf{p})}\right]^\mr{T}
           \,\mbf{V_\dEdX}^{\!\!\!-1}\,\left[\,\widehat{\mbf{\dEdX}}-\mbf{f(\mbf{p})}\right]
\eee
This minimization works well if a large fraction of the shower has been observed
below and above the shower maximum. If this is not the case, or even worse, if the shower
maximum is outside the field of view, the problem is under-determined, i.e.~the experimental
information is not sufficient to reconstruct all four Gaisser-Hillas parameters.
This complication can be overcome by weakly constraining $X_0$ and $\lambda$ to their
average values $\langle X_0\rangle$ and $\langle\lambda\rangle$. The new 
minimization function is then the modified $\chi^2$
\bbe
  \chi^2=\chi^2_\mr{GH}+\frac{(X_0-\langle X_0\rangle)^2}{V_{X_0}}
          +\frac{(\lambda-\langle \lambda\rangle)^2}{V_\lambda} \,,
\eee
where the variance of $X_0$ and $\lambda$ around their mean values are in the denominators.\\
In this way, even if $\chi^2_\mr{GH}$ is not sensitive to $X_0$ and $\lambda$, the minimization
will still converge. On the other hand, if the measurements have small statistical uncertainties and/or
cover a wide range in depth, the minimization function is flexible enough to allow for 
shape parameters differing from their mean values. These mean values can be determined 
from air shower simulations or, preferably, from high quality data profiles which can be
reconstructed without constraints.\\

\bibliography{icrc0972}
\bibliographystyle{unsrt}
\end{document}